# EDA-RoF: Elastic Digital-Analog Radio-Over-Fiber (RoF) Modulation and Demodulation Architecture Enabling Seamless Transition Between Analog RoF and Digital RoF

**Xiaobo Zeng[1,*], Pan Liu[1], Liangcai Chen[1], Ruonan Deng[2]**
[1]*Hunan Province Key Laboratory of Credible Intelligent Navigation and Positioning, Key Laboratory of Intelligent, Computing and Information Processing of Ministry of Education, National Center for Applied Mathematics in Hunan, Xiangtan University, Xiangtan, 411105, China.*
[2]*College of Meteorology and Oceanology, National University of Defense Technology, Changsha, 410073, China.*
*\*Corresponding Author: xiaobo.zeng@xtu.edu.cn*

**Abstract:** We propose and demonstrate an elastic digital-analog radio-over-fiber (RoF) modulation and demodulation architecture, seamlessly bridging A-RoF and D-RoF solutions, achieving quasi-linear SNR scaling with respect to $1/\eta$, and evidenced by $R^2$=0.9908. © 2026 The Author(s)

## 1. Introduction

Mobile fronthaul constitutes critical fiber-based component and be tasked with the communication between distributed units (DUs) and remote radio units (RUs) within centralized radio access networks (C-RANs)[1]. To address projected 10- to 100-fold capacity growth[2], radio-over-fiber (RoF) techniques are investigated for optimizing the inherent trade-off between recovered signal-to-noise ratio (SNR) and spectral efficiency (SE)[3]. These techniques are typically classified as digital RoF (D-RoF) and analog RoF (A-RoF)[3]. A typical D-RoF of common public radio interface (CPRI) achieves the recovered SNR exceeding 80 dB by 15-bit quantization[4], while it incurs substantial bandwidth overhead. Conversely, A-RoF offers superior SE yet remains susceptible to link noise, limiting achievable SNR to approximately 25 dB[5]. To effectively achieve a favorable trade-off between SNR and SE, hybrid digital-analog RoF (DA-RoF) scheme have been proposed[6], extending to dual-polarization coherent systems with multi-dimensional encoding, including phase, polarization, and space diversity, to achieve Tb/s CPRI-equivalent rates[7, 8]. Further, a fractional-order digital-analog RoF (FDA-RoF) scheme has been introduced to enable fractional bandwidth ratios and linear SNR scaling between adjacent integer-order DA-RoF signals[5]. However, the seamless transition between A-RoF and D-RoF is still a significant challenge.

In this paper, we propose and demonstrate an elastic digital-analog RoF (EDA-RoF) modulation and demodulation architecture, which can adaptively modulate and demodulate the target signal based solely on the desired SE, achieving a SNR gain that scales quasi-proportionally with SE. The proposed EDA-RoF seamlessly bridges the functional gap between conventional A-RoF and D-RoF solutions while maintaining spectral flexibility. Numerical evaluations are conducted. As the increases of $1/\eta$ from 1 to 5, the recovered SNR of the proposed EDA-RoF system exhibits quasi-inear scaling with $R^2$=0.9908. The proposed EDA-RoF technique provides a viable solution for future flexible and spectrally-efficient fronthaul-like networks.

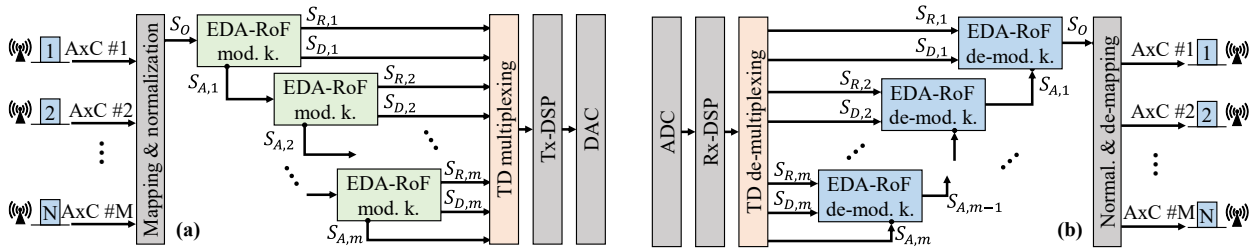

Fig. 1. Schematic diagrams of the general (a) modulation and (b) demodulation of EDA-RoF. TD: time division; mod. k.: modulation kernel; de-mod. k.: demodulation kernel.

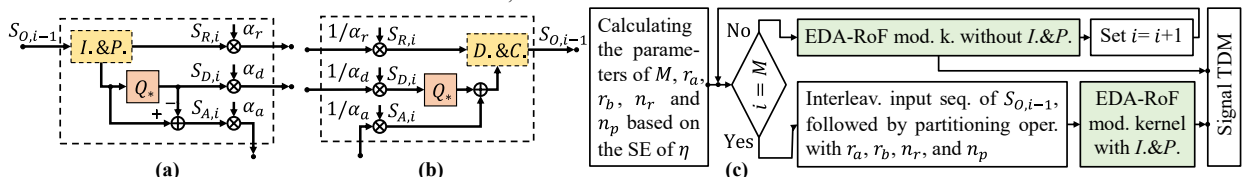

Fig. 2. Schematic diagrams of the general kernels of (a) modulation and (b) demodulation of EDA-RoF. (c) Flow chart of EDA-RoF modulation based on target spectral efficiency. k.: kernel; *I*.: interleaving; *P*.: partitioning; *D*.: de-interleaving; *C*.: combining; *Q*: quantization function.

## 2. Principle

Based on the fundamental differences of quantization between A-RoF and D-RoF, cascaded architecture incorporating a multi-order low-bit-quantization modulation kernel, named EDA-RoF modulation kernel, is proposed, as shown in Fig. 1(a), which integrates the high-fidelity of D-RoF with the spectrally efficient advantage of A-RoF, and seamlessly bridging the operational gap between conventional A-RoF and D-RoF with elastic SE, named elastic digital-analog RoF (EDA-RoF). Specifically, for the given uplink wireless signals $S_O$ aggregated in frequency or chirp domain[9], the $M$-order EDA-RoF generates $M$ D-RoF segments $S_D$, 1 A-RoF segment $S_A$ and 1 residual A-RoF (RA-RoF) segment $S_R$. decomposed signal components are mathematically expressed as

$$S_{R,i}, S_{D,i}, S_{A,i} = F_{EDA-RoF}(S_{O,i-1}, \eta) \quad (1)$$

where $\eta$ denotes the target spectral efficiency relative to A-RoF. $F_{EDA-RoF}(\cdot)$ represents the EDA-RoF modulation kernel with the input signal of $S_{O,i-1}$, as illustrated in Fig. 2(a), where $S_{O,i-1} = S_O$ if $\eta \geq 1/2$ and $S_{O,i-1} = S_{A,i-2}$ otherwise. For the $i$-th order kernel, the input signal $S_{O,i-1}$ is firstly interleaved, and then partitioned into a residual signal $S_{R,i}$ and a transitional signal to be processed, $S_{P,i}$ in time domain. The residual signal $S_{R,i}$ conforms the A-RoF signal, while the $S_{P,i}$ is decomposed into digital ($S_{D,i}$) and analog ($S_{A,i}$) components. The $S_{D,i}$ can be expressed as[10]

$$S_{D,i} = Q_*(S_{P,i}, \mathbf{n}) \quad (2)$$

where $Q_*$ denotes the quantization function, encompassing $Q_P$ for polar coordinates and $Q_C$ for Cartesian coordinates. $\mathbf{n}$ represents the set of quantization factor: $\mathbf{n} = [n_a, n_\varphi]$, if $Q_P$ is applied, with $n_a$ and $n_\varphi$ specifying the target levels of amplitude and phase to be quantized, respectively. For $Q_C$, $\mathbf{n} = [n_a]$, where $n_a$ denotes the target levels of the real and imagination components. The $S_{A,i}$ can be derived by the subtraction operation as [6]

$$S_{A,i} = S_{P,i} - S_{D,i} \quad (3)$$

Then, the three signal of $S_{R,i}$, $S_{D,i}$ and $S_{A,i}$ generated by the EDA-RoF modulation kernel are power-optimized by the respective scaling factors of $\alpha_r$, $\alpha_d$ and $\alpha_a$, respectively. Prior to the model of digital signal processing, these signals of $S_{R,i}$, $S_{D,i}$ and $S_{A,i}$ with $i \in N$ are multiplexed in time domain, as shown in Fig. 1(a).

To achieve elastic SE adjustment, the system parameters of EDA-RoF, including the number of modulation order $M \in N^+$, characteristic factors of $r_a$ and $r_b$, and partitioning factors of $n_p$ and $n_r$, can be derived only based on the target SE. The flow chart of EDA-RoF modulation is illustrated in Fig. 2(c). For an $M$-order EDA-RoF system, the achievable SE is reduced to $1/M \geq \eta \geq 1/(M+1)$. The 1-order EDA-RoF with $\eta = 1$ corresponds to the traditional A-RoF. As $M$ increases large enough and the quantized residual signal becomes negligible, EDA-RoF can be evolved to conventional D-RoF. Thus, the EDA-RoF framework can elastically modulate the aggregated wireless signal with any SE before the fronthaul transmission, facilitating seamless transition between A-RoF and D-RoF. In addition, the demodulation of EDA-ROF is exactly the opposite process of above modulation, as shown in Fig. 1(b) and Fig. 2(b).

## 3. Simulation Setup and Results

The simulation setup is depicted in Fig. 3(a), where a single-polarization coherent system with optical back-to-back (OB2B) connection is modeled. The frequency response of transmitter measured from our experimental setup is incorporated into the model, as shown in Fig. 3(d). Based on the datasheet of Keysight M8194A arbitrary waveform generator, the digital-to-analog converter (DAC) and analog-to-digital converter (ADC) are modeled as ideal quantizer with an effective number of bits (ENOB) of 6 [11]. The electric amplifiers (EAs) are characterized using the Rapp model of solid-state power amplifier as[12], In order to focus on the elastic SE, the back-off value is set to 9 as[10] for the weak nonlinearity. The amplified signals are then modulated onto an optical carrier via a Mach-Zehnder interference based single-polarization in-phase-quadrature (IQ) modulator with a half-wave voltage $V_\pi$ = 4V. At the receiver side, a variable optical attenuator (VOA) is used to adjust the received optical power (ROP) of target signal, followed by a single-polarization coherent receiver with balanced photodiodes. The responsivity, dark current, and the power spectral density of thermal noise of used photodiodes are defined as 0.8 A/W, $5 \times 10^{-9}$ A and $10 \times 10^{-12}$ A/Hz$^{1/2}$, respectively. The laser and local oscillator (LO) operate as continuous-wave sources with the central wavelength of 1550 nm, the linewidth of 100 kHz, and the relative intensity noise factor of -150 dBc/Hz.

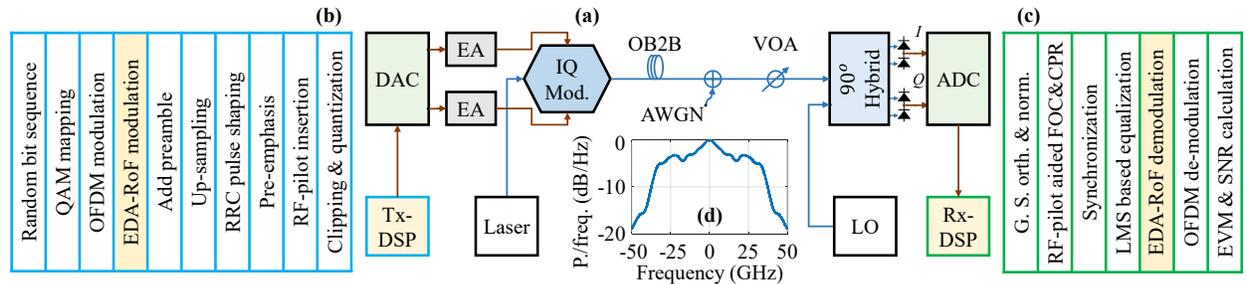

Fig. 3. (a) Simulation setup with optical back-to-back connection and (b) Tx-DSP and (c) Rx-DSP. (d) Frequency response measured from our experimental setup. EA: electric amplifier; LO: local oscillator; FOC: frequency offset correction; CPR: carrier phase recovery.

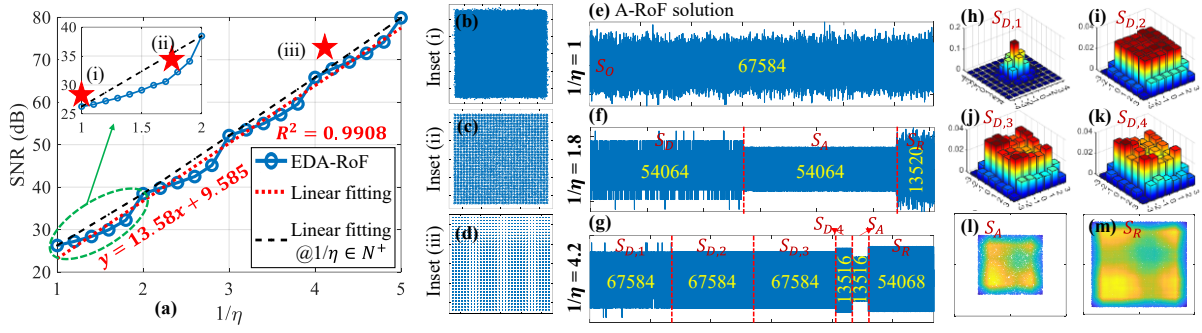

Fig. 4. (a) Measured SNR of the recovered wireless signals as the function of $1/\eta$. (b)-(d) Inset (i)-(iii) present the constellations of the recovered wireless signals of EDA-RoF system with $1/\eta$ = 1, 1.8, and 4.2, respectively. (e)-(g) illustrate the waveforms of the EDA-RoF modulated signals with $1/\eta$ = 1, 1.8, and 4.2, respectively. (h)-(m) Constellations of D-RoF segments of $S_{D,1}$, $S_{D,2}$, $S_{D,3}$, $S_{D,4}$, A-RoF segment of $S_A$ and residual A-RoF segment of $S_R$.

For the digital signal processing models at the transmitter side (Tx-DSP), as shown in Fig. 3(b), the generated random bit sequence is mapped into symbols with the format of 1024-ary quadrature amplitude modulation (QAM) and the length of $2^{16}$. The symbol sequence is paralleled into $2^{10}$ columns, and to emulate the received wireless signals of $x(1)$, $x(2)$, ..., and $x(M)$ as shown in Fig. 1(a). Then a $2^{10}$-point inverse fast Fourier transformation (IFFT) is employed for OFDM modulation. The EDA-RoF modulation is conducted as shown in Fig. 1(a) and Fig. 2(a). The modulated signal is encapsulated with a preamble including 64 symbols for synchronization and the training of equalizer in the Rx-DSP. Up-sampling with 2 samples-per-symbol (sps) and a root-raised cosine filter (RRC) based pulse shaping with a roll-off factor of 0.1 are used. The shaped signal with the bandwidth of 38.5 GHz. The digital pre-emphasis based on zero-forcing is used to pre-compensate the bandwidth limitation of the transmitter. A RF pilot tone is inserted to facilitate the frequency offset correction (FOC), and carrier phase recovery (CPR). After clipping and quantization, the signals are loaded into the DAC, where the clipping ratio is optimized. The digital signal processing models at the receiver side (Rx-DSP) including Gram-Schmidt orthogonalization and normalization, RF-pilot aided FOC & CPR, synchronization, least-mean-squares (LMS) based equalization, EDA-RoF demodulation, OFDM demodulation, and EVM & SNR calculation are used.

In the simulation, the parameters of $n$, and scaling factors of $\alpha_r$, $\alpha_d$ and $\alpha_a$ are optimized. The measured SNR of the recovered wireless signals as the function of $1/\eta$ are shown in Fig. 4(a), where the variation range of $1/\eta$ is from 1 to 5 with a step size of 0.1. The proposed EDA-RoF technique seamlessly integrates with A-RoF. The recovered SNR demonstrates a quasi-linear scaling with the increases of $1/\eta$ and with $R^2 = 0.9908$. Additionally, with the increase of $1/\eta$, the $M$ becomes large enough, and the quantized residual signals of $S_A$ and $S_R$ in the EDA-RoF become negligible. Consequently $S_A$ and $S_R$ can be eliminated from the time domain multiplexed signal, and without degrading the recovered SNR. Numerical results show that EDA-RoF with $M = 7$ and without $S_A$ and $S_R$ achieve a SNR of 83.281 dB, evolving into the D-RoF through multi-stage cascaded architecture with low-resolution quantizers.

## 4. Conclusions

This paper proposes and demonstrates an EDA-RoF technique which enables adaptive modulation and demodulation based solely on the target SE, achieving a quasi-proportional scaling of SNR gain with respect to $1/\eta$ and with $R^2 = 0.9908$. The proposed EDA-RoF technique seamlessly bridges the gap between conventional A-RoF and D-RoF architectures, offering a viable solution for future flexible and spectrally-efficient fronthaul-like networks.

## 5. Acknowledgements
This work was supported by Doctoral Research Startup Foundation of Xiangtan University under Grant 24QDZ32.